\documentstyle[11pt]{article}

\def\d{{\partial}}
\newcommand{\be}{\begin{equation}}
\newcommand{\ee}{\end{equation}}
\newcommand{\ba}{\begin{eqnarray}}
\newcommand{\ea}{\end{eqnarray}}
\newcommand{\no}{\nonumber}
\newcommand{\al}{\alpha}
\newcommand{\bt}{\beta}
\newcommand{\ga}{\gamma}

\newcommand{\la}{\langle}
\newcommand{\ra}{\rangle}
\newcommand{\Op}{{\cal O}}

\newcommand{\h}{\hskip1mm}

\begin{document}

\thispagestyle{empty}

\begin{flushright}
UT-802 \\
YITP-02-98\\
January, 1998 \\
\end{flushright}

\bigskip

\begin{center}
{\Large \bf Quantum Cohomology
at Higher Genus: Topological Recursion Relations and Virasoro Conditions}
\end{center}
\bigskip

\vskip25mm

\begin{center}

Tohru Eguchi 

\bigskip

{\it Department of Physics, Faculty of Science,

\medskip

University of Tokyo, 

\medskip

Tokyo 113, Japan}

\bigskip

\bigskip

and

\bigskip

\bigskip

Chuan-Sheng Xiong

\medskip

{\it Yukawa Institute for Theoretical Physics,

\medskip

Kyoto University 

\medskip

Kyoto 606, Japan}
\end{center}

\bigskip

\vskip3cm

\begin{abstract}
We construct topological recursion relations (TRR's) at higher genera $g\ge2$
for general 2-dimensional topological field theories coupled to
gravity. These TRR's when combined with Virasoro conditions enable one
to determine the number of higher genus holomorphic curves 
in any Fano varieties. In the case of $CP^2$ we reproduce the known
results at genus $g=2$.
\end{abstract}

\newpage
\pagenumbering{arabic}

\noindent{\large \bf 1. Introduction} \\

\bigskip
Recently we have proposed that the partition functions of
topological string theories compactified on an arbitrary K\"{a}hler
manifold 
are highest-weight states of a Virasoro algebra \cite{EHX,EJX,Ka}. We
have shown that 
the Virasoro conditions together with topological recursion relations
reproduce known instanton numbers (numbers of holomorphic curves)
at genus 0 and 1 of various Fano varieties. 
Topological recursion relations (TRR's) exist at genus $g=0$ and $g=1$
\cite{Wittena} and convert correlation functions
of gravitational descendants into those of primary fields which may be
evaluated directly by the method of algebraic geometry.
In the case of higher-genera $g\ge 2$, however, no
TRR have been so far available and it was not possible to compare the 
predictions of
Virasoro conditions with the known geometrical data on the higher genus 
curves.

Recently Getzler has announced the existence
of a TRR at genus $g=2$ \cite{Getzlera} which involves 
descendants of 
degree $n,\hskip1mm
n-1$ and $n-2$. His analysis is based on the study of a
linear relation among homology cycles on the moduli space of
holomorphic maps. A precise form of his formula, however, is not yet available.

In this paper we propose a somewhat different form of TRR at higher genera 
starting from a simple assumption on the
dependence of higher-genus free energies on genus=0 primary correlation
functions. Our recursion relation at genus $g$ involves gravitational
descendants of degree $n,\h n-1,\cdots, n-3g+1$. In the case of genus 
$2$ descendants of degree $n,\h n-1,\cdots,n-5$ appear and thus
our relation is somewhat weaker than that of Getzler's. Nevertheless, these
TRR's can be used together with the Virasoro conditions in order  
to completely determine the number of holomorphic curves of arbitrary degree and
genus. In the case of $CP^2$ at genus $g=2$ we explicitly verify that our TRR and
Virasoro conditions reproduce the known results of \cite{Vai,DI,CH}. 

\bigskip

\bigskip

\noindent{\large \bf 
2. Topological Recursion Relation at Higher Genus} \\

Our basic assumption is that the genus-$g$ free energy of topological
string theory is a function depending only on the primary multi-point functions of
genus $g=0$. Let us consider the case of a theory
with primary fields $\{\Op_{\al}\} (\al=0,1\cdots,N)$ coupled to
the perturbation parameters
$\{t^{\al}\}$. Gravitational descendants and their
couplings are denoted as
$\{\sigma_n(\Op_{\al})\}$ and $\{t^{\al}_{n}\}$ $ (n=0,1,2,\cdots)$,
respectively. 
We define genus=0 
correlation functions as $u_{\al_1\al_2\cdots\al_j}\equiv\la P
\Op_{\al_1}\Op_{\al_2}\cdots \Op_{\al_j}\ra_0$ where $P$ denotes the
puncture operator. Our assumption is that the genus-$g$ free
energy $F_g$ is a function only of
the genus=0 correlation functions
$u_{\al_1},u_{\al_1\al_2},\cdots,u_{\al_1\al_2\cdots\al_{3g-1}}$
\be
F_g(t)=F_g(u_{\al_1}(t),u_{\al_1\al_2}(t),\cdots,
u_{\al_1\al_2\cdots\al_{3g-1}}(t)), \hskip25mm g\ge 1.
\ee
Here $t$ stands for all the couplings $\{t^{\al}_n\}$ of the large phase space.
Note that the dependence on the parameters $t$ of the free energy $F_g(t)$ 
occurs only
through the functions $u_{\al_1\al_2\cdots\al_{j}}(t) \h (j=1,2,\cdots,3g-1)$.

Equation (1) is known to hold in the 2-dimensional pure gravity
theory.
For instance, the genus 1,2 and 3 free energies are given by \cite{DW,IZ}
\ba
&&F_1={1 \over 24}\log u', \hskip10mm
F_2={u''^3 \over 360u'^4}-{7u''u^{(3)} \over 1920u'^3}+{u^{(4)} \over
1152u'^2}, \\
&&F_3=
   -{{5 {{u''}^6}}\over {648 {{u'}^8}}} + 
   {{59 {{u''}^4} u^{(3)}}\over {3024 {{u'}^7}}} - 
   {{83 {{u''}^2} {{u^{(3)}}^2}}\over {7168 {{u'}^6}}} + 
   {{59 {{u^{(3)}}^3}}\over {64512 {{u'}^5}}}
    -{{83 {{u''}^3} u^{(4)}}\over {15120 {{u'}^6}}} + 
   {{1273 u'' u^{(3)} u^{(4)}}\over {322560 {{u'}^5}}} \no \\
&&\hskip15mm-{{103 {{u^{(4)}}^2}}\over {483840 {{u'}^4}}} + 
   {{353 {{u''}^2} u^{(5)}}\over {322560 {{u'}^5}}} 
   -{{53 u^{(3)} u^{(5)}}\over {161280 {{u'}^4}}} - 
   {{7 u'' u^{(6)}}\over {46080 {{u'}^4}}} + 
   {{u^{(7)}}\over {82944 {{u'}^3}}}
\ea
where $u=\la PP\ra_0$ and $'$ denotes the $t^0$-derivative.
Eq.(1) is also known to hold 
in some cases of the 2-dimensional gravity coupled to minimal matter
at lower genera \cite{EYY}. It is also valid
in the case of $CP^1$ model \cite{EY}. 
Eq.(1) means that higher genus amplitudes are expressed in terms of the
genus=0 data and suggests a possible reinterpretation of the
world-sheet topological theory as a field
theory on the target space \cite{IZ,EYY}. We now assume that (1) is a universal
feature of 2-dimensional topological field theories coupled to gravity. 
 
It is then easy to derive our TRR's. Let us first consider for 
simplicity the case of genus=1. Genus-1 free energy depends on
$\{u_{\al}\}$ and $\{u_{\al\bt}\}$. We then have
\ba
&&{\partial F_1 \over \partial t^{\al}_n}={\partial u_{\mu}
\over \partial t^{\al}_n}{\partial F_1 \over \partial u_{\mu}}+{\partial
u_{\mu\nu} \over \partial t^{\al}_n}{\partial F_1 \over \partial u_{\mu\nu}}
=\la \sigma_n(\Op_{\al})\Op_{\mu}P\ra_0{\partial F_1 \over
\partial u_{\mu}}+{\partial \la \sigma_n(\Op_{\al})\Op_{\nu}P\ra_0
\over \partial t_{\mu}}{\partial F_1 \over
\partial u_{\mu\nu}}. \no \\
\label{TRR1}&&
\ea
At $n=0$ eq.(\ref{TRR1}) becomes
\be
\la \Op_{\al}\ra_1=\la\Op^{\al}\Op_{\mu}P\ra_0{\partial F_1 \over
\partial u_{\mu}}+\la \Op^{\al}\Op_{\mu}\Op_{\nu}P\ra_0{\partial F_1 \over
\partial u_{\mu\nu}}.
\label{TRR2}
\ee
We use the genus=0 TRR
\be
\la \sigma_n(\Op_{\al})AB\ra_0=\la \sigma_{n-1}(\Op_{\al})
\Op_{\gamma}\ra_0\la\Op^{\gamma}AB\ra_0
\label{TRRg0}\ee
to rewrite (\ref{TRR1}) as
\ba
&&\la \sigma_n(\Op_{\al})\ra_1=\la
\sigma_{n-1}(\Op_{\al})\Op_{\bt}\ra_0\la \Op^{\bt}\Op_{\mu}P\ra_0{\partial F_1 \over
\partial u_{\mu}}\no \\
&&+\Big(\la \sigma_{n-1}(\Op_{\al})\Op_{\bt}\ra_0\la\Op^{\bt}\Op_{\mu}\Op_{\nu}P\ra_0
+\la \sigma_{n-1}(\Op_{\al})\Op_{\mu}\Op_{\bt}\ra_0\la\Op^{\bt}\Op_{\nu}P\ra_0\Big)
{\partial F_1 \over
\partial u_{\mu\nu}}.
\label{TRR3}\ea
Putting $n=1$ in (\ref{TRR3}) gives
\ba
&&\la \sigma_1(\Op_{\al})\ra_1=\la
\Op_{\al}\Op_{\bt}\ra_0\la \Op^{\bt}\Op_{\mu}P\ra_0{\partial F_1 \over
\partial u_{\mu}}\no \\
&&+\Big(\la \Op_{\al}\Op_{\bt}\ra_0\la\Op^{\bt}\Op_{\mu}\Op_{\nu}P\ra_0
+\la \Op_{\al}\Op_{\mu}\Op_{\bt}\ra_0\la\Op^{\bt}\Op_{\nu}P\ra_0\Big)
{\partial F_1 \over \partial u_{\mu\nu}} \no \\
&&=\la \Op_{\al}\Op_{\bt}\ra_0 \la\Op^{\bt}\ra_1+
\la \Op_{\al}\Op_{\mu}\Op_{\bt}\ra_0\la\Op^{\bt}\Op_{\nu}P\ra_0
{\partial F_1 \over \partial u_{\mu\nu}}
\label{TRR4}\ea
where (\ref{TRR2}) has been used.
By making use of (\ref{TRR2}) and (\ref{TRR4}) eq.(\ref{TRR3}) is then reexpressed as 
\be
\la \sigma_n(\Op_{\al})\ra_1=\la
\sigma_{n-1}(\Op_{\al})\Op_{\bt}\ra_0\la\Op^{\bt}\ra_1+\la
\sigma_{n-2}(\Op_{\al})\Op_{\gamma}\ra_0
\Big(\la \sigma_1(\Op^{\gamma})\ra_1-\la \Op^{\gamma}\Op_{\bt}\ra_0 \la\Op^{\bt}\ra_1\Big)
\label{TRR5}\ee
Eq.(\ref{TRR5}) is our TRR at genus=1. It appears somewhat different from the
standard TRR \cite{Wittena}
\be
\la\sigma_n(\Op_{\al})\ra_1={1 \over 24}\la \sigma_{n-1}(\Op_{\al})\Op_{\bt}\Op^{\bt}\ra_0
+\la \sigma_{n-1}(\Op_{\al})\Op_{\bt}\ra_0\la\Op^{\bt}\ra_1.
\label{TRRg1}\ee
However, when one uses the structure of the genus=1 free energy
\be
F_1={1 \over 24}\log\det (u_{\al\bt})+f_1(u_{\al})
\label{feg1}\ee
one may easily check (\ref{TRR5}) and (\ref{TRRg1}) are equivalent.

By repeating the same procedure as above we can derive the TRR at
genus=2
\ba
&&\la\sigma_{n+5}(\Op_{\al})\ra_2=\la\sigma_{n+4}(\Op_{\al})\Op_{\bt}\ra_0
A_0^{\bt}
+\la\sigma_{n+3}(\Op_{\al})\Op_{\bt}\ra_0 A_1^{\bt}
+\la\sigma_{n+2}(\Op_{\al})\Op_{\bt}\ra_0 A_2^{\bt} \no \\
&&\hskip40mm +\la\sigma_{n+1}(\Op_{\al})\Op_{\bt}\ra_0 A_3^{\bt}
+\la\sigma_{n}(\Op_{\al})\Op_{\bt}\ra_0 A_4^{\bt} 
\label{TRRg2}\ea
where
\ba
&&A_0^{\bt}\equiv\la \Op^{\bt}\ra_2  \\
&&A_1^{\bt}\equiv\la \sigma_1(\Op^{\bt})\ra_2
-\la \Op^{\bt}\Op_{\ga}\ra_0 A_0^{\ga}
 \\
&&A_2^{\bt}\equiv\la \sigma_2(\Op^{\bt})\ra_2
-\la \sigma_1(\Op^{\bt})\Op_{\ga}\ra_0
A_0^{\ga}-\la \Op^{\bt}\Op_{\ga}\ra_0\cdot A_1^{\ga} \\
&&A_3^{\bt}\equiv\la \sigma_3(\Op^{\bt})\ra_2
-\la \sigma_2(\Op^{\bt})\Op_{\ga}\ra_0
A_0^{\ga}-\la \sigma_1(\Op^{\bt})\Op_{\ga}\ra_0\cdot A_1^{\ga}
-\la \Op^{\bt}\Op_{\ga}\ra_0 \cdot A_2^{\ga} \\
&&A_4^{\bt}\equiv\la \sigma_4(\Op^{\bt})\ra_2
-\la \sigma_3(\Op^{\bt})\Op_{\ga}\ra_0
A_0^{\ga}-\la \sigma_2(\Op^{\bt})\Op_{\ga}\ra_0\cdot A_1^{\ga}
-\la \sigma_1(\Op^{\bt})\Op_{\ga}\ra_0 \cdot A_2^{\ga} \no \\
&&-\la \Op^{\bt}\Op_{\ga}\ra_0\cdot A_3^{\ga}.
\ea
Thus all the descendants $\{\sigma_n(\Op_{\al}), n\ge 5\}$ may be eliminated
in favor of $\{\sigma_i(\Op_{\al}), i=1,2,3,4\}$ at genus
$g=2$. An explicit form of the above TRR is presented in the Appendix.

Similarly, TRR's at arbitrary genus ($g\ge 1$) are given by
\ba
&&\la\sigma_{n+3g-1}(\Op_{\al})\ra_g=
\sum_{j=0}^{3g-2}\la\sigma_{n+3g-2-j}(\Op_{\al}\Op_{\bt})\ra_0 A_j^{\bt},  \\
&&A_0^{\bt}\equiv \la\Op^{\bt}\ra_g,  \\
&&A_{j}^{\bt}=\la\sigma_{j}(\Op^{\bt})\ra_g-\sum_{k=0}^{j-1}
\la\sigma_k(\Op^{\bt})\Op^{\ga}\ra_0A_{j-k-1}^{\ga}.
\ea
Thus the descendant degrees are reduced to $n\le 3g-2$ 
at genus $g$. These TRR are not quite as efficient as the standard
TRR at genus $0$ and $1$ which reduce the descendant degrees all the way 
to zero. As we shall see, however, they are powerful enough to 
determine instanton numbers of arbitrary genera when combined together with
the Virasoro conditions.

\bigskip

\bigskip

\noindent{\large \bf 
3. Virasoro Conditions and Higher Genus Curves in $CP^2$} \\

We now turn to the application of our TRR. In order to fix
our discussions let us consider the case of 
$CP^2$ and determine the number of its genus=2 curves.

Let us first recall the Virasoro conditions for $CP^2$ \cite{EHX}
\be
L_n Z=0, \hskip30mm n \ge -1
\label{Vcondition}\ee
where
\ba
&&L_{-1}=\sum_{\al=0}^2\sum_{m=1}^{\infty}t^{\al}_m\partial_{m-1,\al}
+{1 \over 2 \lambda^2}\sum_{\al=0}^2t^{\al}t_{\al}, \\
&&\hskip-8mm L_0=\sum_{\alpha=0}^2 \sum_{m=0}^\infty 
(b_\alpha+m)t_m^\alpha\d_{m,\alpha}
+3\sum_{\alpha=0}^1 \sum_{m=0}^\infty t_m^\alpha\d_{m-1,\alpha+1}
+\frac{1}{2\lambda^2}\sum_{\alpha=0}^{1} 3t^\alpha
t_{\alpha+1}-\frac{5}{16}, \no \\
&& \\
&&\hskip-8mm L_n=\sum_{m=0}^\infty\sum_{\alpha=0}^2
\sum_{j=0}^{2-\alpha}
3^jC_\alpha^{(j)}(m,n) t_m^\alpha\d_{m+n-j,\alpha+j} \hskip20mm n\ge 1
\label{ln} \no \\
&&\hskip-5mm+\frac{\lambda^2}{2}\sum_{\alpha=0}^2
\sum_{j=0}^{2-\alpha}\sum_{m=0}^{n-j-1}
3^jD_\alpha^{(j)}(m,n) \d^\alpha_m\d_{n-m-j-1,\alpha+j}
+\frac{1}{2\lambda^2}\sum_{\alpha=0}^{1-n} 
3^{n+1}t^\alpha t_{\alpha+n+1}.
\ea
Here $b^0=-1/2,b^1=1/2,b^2=3/2$ and $b_{\al}=1-b^{\al}$ and
the coefficient functions are defined by
\ba
&&C_\alpha^{(j)}(m,n)\equiv
\frac{\Gamma(b_\alpha+m+n+1)}{\Gamma(b_\alpha+m)}
\sum_{m\le \ell_1<\ell_2<\cdots<\ell_j\le m+n}
\bigg(\prod_{i=1}^j \frac{1}{b_\alpha+\ell_i}\bigg), 
\label{cf}\\
&& \no \\
&&\hskip-5mmD_\alpha^{(j)}(m,n)\equiv 
\frac{\Gamma(b^\alpha+m+1)\,\Gamma(b_\alpha+n-m)}
{\Gamma(b^\alpha)\,\,\Gamma(b_\alpha)}\times\sum_{-m-1\le \ell_1<\ell_2<\cdots<\ell_j\le n-m-1}
\bigg(\prod_{i=1}^j \frac{1}{b_\alpha+\ell_i}\bigg). \no \\
&&
\label{df}\ea
Conventional
notations are $t^0\equiv t^P,t^1\equiv t^Q$ and $t^2\equiv t^R$. 

A parameter $\lambda$ denotes the string coupling constant and the free energy has 
the genus expansion
\be
Z=\log F, \hskip20mm F=\sum_{g=0}\lambda^{2g-2}F_g.
\ee
In the small phase space ($t^{\al}_n=0, n>0$ except $t^P_1=-1$) the
genus-g free energy has a structure
\be
F_g=F^{cl}_g+\sum_{d=1}{N^g_d \over (3d+g-1)!} (t^R)^{3d+g-1}e^{dt^Q}
\label{instexp}\ee
where 
$N^g_d$ denotes the number of degree $d$, genus $g$ irreducible curves passing through
$3d+g-1$ points in $CP^2$. The classical part of the free energy 
$F^{cl}_g$ is non-vanishing only at $g=0$ and $1$,
$F^{cl}_0=
t^P(t^Q)^2/2+(t^P)^2t^R/2, F^{cl}_1=-t^Q/8$.

In order to determine the instanton expansion (\ref{instexp}) for $g=2$ free
energy, 
we first have to determine the
values of the ``initial'' descendants 
$\la\sigma_i(\Op_{\al})\ra_2 \h (i=1,2,3,4)$. We consider the Virasoro
conditions (\ref{Vcondition}) for $n=1,2\cdots,12$ in the small phase space 
and substitute into these equations 
the known values of the correlation functions at genus $0,1$.
We also use TRR at $g=2$ (\ref{TRRg2}) to rewrite one-point functions of higher
descendants in terms of those of initial descendants.
Then the Virasoro conditions provide 12
linear relations for 12
unknowns $\la\sigma_i(\Op_{\al})\ra_2 \h (\al=0,1,2,\h i=1,2,3,4)$ and one can
determine them order by order in the instanton expansion.  

If one further considers the next Virasoro condition $L_n$ with $n=13$ and
substitutes into it the determined values of the initial descendants, it then
gives a prediction on the $g=2$ free energy.
One finds upto degree 10
\ba
F_2(t^Q,t^R)&=&\frac{27}{13!}\,(t^R)^{13}\, e^{4t^Q}+\frac{36855}{16!}\,(t^R)^{16}\,e^{5t^Q}
+\frac{58444767}{19!}\,(t^R)^{19}\, e^{6t^Q} \no \\ \no \\
&+&\frac{122824720116}{22!}\,(t^R)^{22}\, e^{7t^Q}
+\frac{346860150644700}{25!}\,(t^R)^{25}\, e^{8t^Q}  \label{feg2}\\ \no \\
&+&\frac{1301798459308709880}{28!}\,(t^R)^{28}\, e^{9t^Q}
+\frac{6383405726993645784000}{31!}\,(t^R)^{31}\, e^{10t^Q}. \no
\ea
First 3 terms in the RHS of (\ref{feg2}) agree with those of Caporaso and
Harris \cite{CH} (Caporaso-Harris gives the number of curves which
contains reducible components. A convenient method of subtraction of
reducible part is described in \cite{Getzlerb}). The rest is our predictions.
One can check higher Virasoro conditions $L_n$ $n>13$ are 
satisfied simultaneously by (\ref{feg2}).

\bigskip

\bigskip

\noindent{\large \bf Discussions}

\bigskip

\bigskip

Precise agreement of our predictions for the genus 2 free
energies of $CP^2$ with the known geometrical data gives a strong support
for the validity of our Virasoro conditions and TRR's.
It now seems that by making use of these equations one may be
able to compute in principle the number of curves of any genus and degree in arbitrary
Fano varieties. For instance, the genus-2 free energy of $CP^3$ upto
degree 3 is computed as
\ba
&&F_2(t^Q,t^R,t^S)=-\frac{1}{288}
+\frac{1}{360}\frac{(t^S)^2}{2!} e^{t^Q}
+\frac{1}{360}\frac{(t^R)^2(t^S)^1}{2!\cdot1!} e^{t^Q}
+\frac{1}{180}\frac{(t^R)^4}{4!} e^{t^Q}\no\\
&&+\frac{7}{240}\frac{(t^R)^2(t^S)^3}{2!\cdot3!} e^{2t^Q}
+\frac{7}{60}\frac{(t^R)^4(t^S)^2}{4!\cdot2!} e^{2t^Q}
+\frac{21}{40}\frac{(t^R)^6(t^S)^1}{6!\cdot1!} e^{2t^Q}
+\frac{161}{60}\frac{(t^R)^8}{8!} e^{2t^Q}\no\\
&&+\frac{1}{12}\frac{(t^S)^6}{6!} e^{3t^Q}
+\frac{5}{12}\frac{(t^R)^2(t^S)^5}{2!\cdot5!} e^{3t^Q}
+\frac{5}{2}\frac{(t^R)^4(t^S)^4}{4!\cdot4!} e^{3t^Q}
+\frac{46}{3}\frac{(t^R)^6(t^S)^3}{6!\cdot3!} e^{3t^Q}  \\
&&+\frac{307}{3}\frac{(t^R)^8(t^S)^2}{8!\cdot2!} e^{3t^Q}
+747\frac{(t^R)^{10}(t^S)^1}{10!\cdot1!} e^{3t^Q}
+5930\frac{(t^R)^{12}}{12!} e^{3t^Q}.\no
\ea
Here the variables $t^Q,t^R,t^S$ are those dual to the
K\"{a}ler class $\omega$ and $\omega^2,\omega^3$ of $CP^3$, respectively.
Number of genus=3 curves in $CP^2$
is currently under study.

Eq.(1) for the genus $g$ free energy is a deep result of the
2-dimensional gravity theory. One may imagine having a set of
space-time fields $\{\phi^{\al}\}$ whose propagator is given by
$u_{\al\bt}^{-1}$ and the j-point vertex by $u_{\al_1\al_2\cdots\al_j}$.
It is known that the genus $g$ free energy $F_g$ equals the sum
of the Feynman amplitudes of $g$-loop 
diagrams made of propagators and j-point vertices ($j=3,4,\cdots,3g-1$) \cite{IZ,EYY}.
The number of different vertices increases as the genus is increased
and hence the system has the characteristic of non-polynomial closed string
field theory. It is quite curious such a space-time interpretation
exists behind our TRR.

Virasoro conditions $L_n Z=0 \h (n\ge -1)$ are not independent in the large phase space
since they form the algebra
$[L_n,L_m]=(n-m)L_{n+m}$. In the small phase space, however, they
become all independent and are used to determine unknown correlation
functions.
For the computation of genus 2 curves in $CP^2$ a large number (13) 
of Virasoro constraints
were imposed.
At the moment the logical relationship between TRR's and the
infinite set of Virasoro
conditions is not completely clear. In the case of 2-dimensional pure 
gravity Virasoro
conditions alone completely determined the amplitude. Hence 
they imply TRR's. On the other hand in the case of 2-dimensional gravity 
coupled to minimal matter, additional 
W-algebra conditions were
necessary to completely determine the amplitudes. Thus the
TRR's are independent of
Virasoro conditions. The present case of topological string theories appears
similar to that of the 2-dimensional gravity coupled to matter.  
It is an important issue if there exist analogues of W conditions
in our present problem.

\bigskip

\bigskip

After completing this manuscript we received a new paper by Getzler 
\cite{getzlerc} where
his TRR at genus=2 is presented. It is easy to check that his equation (6)
is consistent with our genus=2 TRR.

\bigskip

\bigskip

\noindent{\large \bf Note Added} \\

We have obtained the genus=3 free energy of $CP^2$ up to degree $10$

\ba
F_3(t^Q,t^R)&=&\frac{1}{14!}\,(t^R)^{14}\, e^{4t^Q}+\frac{7915}{17!}\,
(t^R)^{17}\,e^{5t^Q}
+\frac{34435125}{20!}\,(t^R)^{20}\, e^{6t^Q} \no \\ \no \\
&+&\frac{153796445095}{23!}\,(t^R)^{23}\, e^{7t^Q}
+\frac{800457740515775}{26!}\,(t^R)^{26}\, e^{8t^Q}  \label{feg3}\\ \no \\
&+&\frac{5039930694167991360}{29!}\,(t^R)^{29}\, e^{9t^Q}
+\frac{38747510483053595091600}{32!}\,(t^R)^{32}\, e^{10t^Q}. \no
\ea

First 3 terms of (\ref{feg3}) agree with \cite{CH}.

\bigskip

\bigskip

\noindent{\large \bf Acknowledgment} \\

\bigskip

We would like to thank E. Getzler and S.K. Yang for discussions.
T.E. also would like to thank I. Ciocan-Fontanine, C. Faber, S. Katz,
Z. Qin and Y. Ruan 
for their interests in this work.

\newpage

\begin{center}
{\large \bf Appendix \\
\medskip
Genus $2$ Topological Recursion Relation}
\end{center}

\bigskip

\bigskip

\ba
&&\hskip-20mm\la \sigma_{n+5}(\Op_{\al})\ra_2=\la
\sigma_n(\Op_{\al})\Op^{\bt}\ra_0\la\sigma_4(\Op_{\bt})\ra_2
+\left[\la\sigma_{n+1}(\Op_{\al})\Op^{\bt}\ra_0-\la\sigma_n(\Op_{\al})\Op_{\ga}\ra_0
\la\Op^{\ga}\Op^{\bt}\ra_0\right]\la\sigma_3(\Op_{\bt})\ra_2 \no \\
\no \\
&&\hskip-20mm+\left[\la\sigma_{n+2}(\Op_{\al})\Op^{\bt}\ra_0-\la\sigma_{n+1}(\Op_{\al})\Op_{\ga}\ra_0
\la\Op^{\ga}\Op^{\bt}\ra_0+\la\sigma_n(\Op_{\al})\Op_{\mu}\ra_0
\Big(\la\Op^{\mu}\Op_{\rho}\ra_0\la\Op^{\rho}\Op^{\bt}\ra_0
-\la\sigma_1(\Op^{\mu})\Op^{\bt}\ra_0\Big)
\right]\la\sigma_2(\Op_{\bt})\ra_2 \no \\ \no \\
&&\hskip-20mm+\left[\la\sigma_{n+3}(\Op_{\al})\Op^{\bt}\ra_0-\la\sigma_{n+2}(\Op_{\al})\Op_{\ga}\ra_0
\la\Op^{\ga}\Op^{\bt}\ra_0+\la\sigma_{n+1}(\Op_{\al})\Op_{\mu}\ra_0
\left\{\la\Op^{\mu}\Op_{\rho}\ra_0\la\Op^{\rho}\Op^{\bt}\ra_0
-\la\sigma_1(\Op^{\mu})\Op^{\bt}\ra_0\right\}
\right.\no \\ \no \\
&&\hskip-20mm
+\la\sigma_n(\Op_{\al})\Op_{\mu}\ra_0\left(-\la\sigma_2(\Op^{\mu})\Op^{\bt}\ra_0+
\la\sigma_1(\Op^{\mu})
\Op_{\rho}\ra_0\la\Op^{\rho}\Op^{\bt}\ra_0\right. \no \\
\no \\
&&\hskip45mm\left.\left.
+\la\Op^{\mu}\Op_{\rho}\ra_0\la\sigma_1(\Op^{\rho})\Op^{\bt}\ra_0
-\la\Op^{\mu}\Op_{\rho}\ra_0\la\Op^{\rho}\Op_{\sigma}\ra_0
\la\Op^{\sigma}\Op^{\bt}\ra_0\right)
\right]\la\sigma_1(\Op_{\bt})\ra_2 \no \\ \no \\
&&\hskip-20mm +\left[\la\sigma_{n+4}(\Op_{\al})\Op^{\bt}\ra_0
-\la\sigma_{n+3}(\Op_{\al})\Op_{\ga}\ra_0
\la\Op^{\ga}\Op^{\bt}\ra_0+\la\sigma_{n+2}(\Op_{\al})\Op_{\mu}\ra_0
\Big(\la\Op^{\mu}\Op_{\rho}\ra_0\la\Op^{\rho}\Op^{\bt}\ra_0
-\la\sigma_1(\Op^{\mu})\Op^{\bt}\ra_0\Big) \right. \no \\ \no \\
&&\hskip-20mm +\la\sigma_{n+1}(\Op_{\al})\Op_{\mu}\ra_0
\left\{-\la\sigma_2(\Op^{\mu})\Op^{\bt}\ra_0+
\la\sigma_1(\Op^{\mu})\Op_{\rho}\ra_0\la\Op^{\rho}\Op^{\bt}\ra_0
\right. \no \\ \no \\
&&\hskip45mm \left.+\la\Op^{\mu}\Op_{\rho}\ra_0\la\sigma_1(\Op^{\rho})\Op^{\bt}\ra_0 
-\la\Op^{\mu}\Op_{\rho}\ra_0\la\Op^{\rho}\Op_{\sigma}\ra_0
\la\Op^{\sigma}\Op^{\bt}\ra_0\right\} \no \\ \no \\
&&\hskip-20mm
+\la\sigma_n(\Op_{\al})\Op_{\mu}\ra_0
\left(-\la\sigma_3(\Op^{\mu})\Op^{\bt}\ra_0
+\la\sigma_2(\Op^{\mu})\Op_{\rho}\ra_0\la\Op^{\rho}\Op^{\bt}\ra_0
+\la\sigma_1(\Op^{\mu})\Op_{\rho}\ra_0\la\sigma_1(\Op^{\rho})\Op^{\bt}\ra_0 
\right.\no \\ \no \\
&&-\la\sigma_1(\Op^{\mu})\Op_{\ga}\ra_0\la\Op^{\ga}\Op_{\rho}\ra_0
\la\Op^{\rho}\Op^{\bt}\ra_0
+\la\Op^{\mu}\Op_{\rho}\ra_0\la\sigma_2(\Op^{\rho})\Op^{\bt}\ra_0
-\la\Op^{\mu}\Op_{\ga}\ra_0\la\sigma_1(\Op^{\ga})\Op_{\rho}\ra_0\la\Op^{\rho}\Op^{\bt}\ra_0
\no \\ \no \\
&&\left.\left.
-\la\Op^{\mu}\Op_{\ga}\ra_0\la\Op^{\ga}\Op_{\rho}\ra_0\la\sigma_1(\Op^{\rho})\Op^{\bt}\ra_0
+\la\Op^{\mu}\Op_{\nu}\ra_0\la\Op^{\nu}\Op_{\rho}\ra_0\la\Op^{\rho}\Op_{\ga}\ra_0
\la\Op^{\ga}\Op^{\bt}\ra_0\right)\right]\la\Op_{\bt}\ra_2 \no
\ea


\end{document}